\documentclass[twocolumn,preprintnumbers,amsmath,amssymb,pre,aps]{revtex4}

\usepackage{braket}
\usepackage{lineno}
\usepackage{comment}
\usepackage[dvipdfmx]{graphicx}
\usepackage{dcolumn}
\usepackage{bm}
\usepackage{comment}
\usepackage[dvipdfmx]{color}
\usepackage{mathrsfs}

\begin{document}

\title{Oscillation collapse in coupled quantum van der Pol oscillators}

\author{Kenta Ishibashi and Rina Kanamoto}
\affiliation{
Department of Physics, Meiji University, Kawasaki, Kanagawa 214-8571, Japan
}
\date{\today}

\begin{abstract}
The classical self-oscillations can collapse merely due to their mutual couplings. 
We investigate this oscillation collapse in quantum van der Pol oscillators.  
For a pair of quantum oscillators, the steady-state mean phonon number is shown to be lower than in the corresponding classical model with a Gaussian white noise that mimics quantum noise. 
We further show within the mean-field theory that a number of globally coupled oscillators undergo a transition from the synchronized periodic motion to the collective oscillation collapse. 
A quantum many-body simulation suggests that the increase in the number of oscillators leads 
to a lower steady-state mean phonon number, bounded below by the mean-field result. 
\end{abstract}

\maketitle

\section{Introduction}

Coupled nonlinear oscillator dynamics is ubiquitously found in nature ranging from biological systems and chemical reactors to mechanical oscillators like pendulums~\cite{Nicolis}.  
In the absence of mutual coupling, the individual oscillator exhibits a stable self-oscillation when the energy gain and loss balance 
by virtue of the nonlinearity. 
When self-sustained oscillators are coupled, they show rich varieties of amplitude and phase dynamics. 
The well-known phase dynamics is synchronization~\cite{Pikovsky}, which is also called phase lock or entrainment depending on the robustness of the lock. 
Not only the classical synchronization, but also its quantum-mechanical aspects are of growing interest~\cite{Zhirov, Holmes:2012ew, Mari:2013cz, Lee:2013go, Xu:2013tm, Walter:2014fx, Lee:2014dq, Walter:2014cs, Ameri:2015kj, Jager} because of the recent experimental developments of cavity cooling and control of micro- and nano-scale mechanical oscillators~\cite{Aspelmeyer}. The phase dynamics of the optomechanical arrays has been studied theoretically and experimentally~\cite{Heinrich:2011jd, Zhang:2012ks, Matheny:2014io,Gil-Santos:2017}.

The quantum amplitude dynamics, on the other hand, has been much less elucidated as compared with the phase dynamics. 
Classically, nonlinear coupled oscillators are known to exhibit amplitude death or oscillator death~\cite{SPR:2012, KVK:2013, KBE, Aronson:1990uq, RSJ, Ermentrout:1990} in the presence of mutual coupling, namely, the completely rest state becomes stable merely 
due to the coupling. 
The natural question then arises if the quantum nonlinear oscillators also exhibit a similar amplitude dynamics where 
the periodic motions collapse into a state very close to the quantum ground state, while the intrinsic quantum fluctuations would prevent the complete oscillator death and replace it with the moderate collapse. 
In particular, understanding amplitude dynamics may find potential applications in an alternative cooling of multiple mechanical oscillators instead of the standard cavity optomechanical cooling that exploits the radiation-pressure backaction to reduce the mean phonon number~\cite{Aspelmeyer}. 
In view of the cavity cooling, quantum fluctuations ultimately set a limit on the achievable lowest temperature of the center-of-mass motion of the mechanical oscillators~\cite{Wilson-Rae:2007, Marquardt:2007}. This fact motivates us to study the oscillation collapse in a full quantum-mechanical manner.

In this paper, we investigate the oscillation collapse in coupled van der Pol (vdP) oscillators~\cite{vdP:1922}. 
The vdP model is a prototypical self-sustained oscillator 
that involves energy gain and nonlinear loss, 
of which classical equation of motion is given by $\ddot{x}=-\omega^2 x + G \dot{x}-8\kappa x^2 \dot{x}$, 
where $x$ is the displacement, $\omega$ the intrinsic frequency in the absence of incoming and outgoing energies, 
and $G$, $\kappa$ are proportional to the gain and loss rates, respectively. 
In a weak nonlinear regime, the above equation is reduced to 
a generic amplitude equation $\dot{\alpha}=(-i\omega+G/2-\kappa |\alpha|^2) \alpha$, 
where $\alpha(t)$ is a complex amplitude. 
The vdP oscillator exhibits a self-oscillation of the amplitude $|\alpha_{\rm{ss}}|=\sqrt{G/ (2\kappa)}$, which corresponds to a circle in the classical phase space $({\rm Re}\alpha, {\rm Im}\alpha)$ with the radius $|\alpha_{\rm ss}|$, called a limit cycle. 
The quantum version of this model has been introduced for the studies of quantum synchronization~\cite{Lee:2013go, Walter:2014fx, Lee:2014dq, Walter:2014cs}, where the phase entrainment instead of a strict phase locking is found to survive in the quantum regime. 
The amplitude dynamics in the quantum vdP oscillator has also been discussed in some degree~\cite{Lee:2013go},
but quantitative characterization of the mean phonon number and collective nature are still elusive. 

Here we first address a pair of quantum vdP oscillators with dissipative coupling, and identify the regime where the mean phonon number is significantly reduced with respect to the coupling strength and frequency mismatch. 
Although the steady state of the coupled oscillators is not the absolute quantum ground state due to the intrinsic noise in quantum systems, we found that the mean phonon number is lower than the one of the classical oscillators with Gaussian white noise. 
We further study the collective amplitude dynamics when more than two quantum vdP oscillators are globally coupled. 
The mean-field theory predicts a transition between the synchronized periodic motion and the oscillation collapse.
We also fully solve the many-body master equation up to seven vdP oscillators, and demonstrate that 
the mean phonon number per oscillator decreases down to the mean-field results as the system size increases.

This paper is organized as follows. 
In Sec.~\ref{N2cl}, the classical dynamics of a pair of vdP oscillators is revisited. 
In Sec.~\ref{N2qm}, we discuss the quantum oscillation collapse as well as its signature. 
In Sec.~\ref{N2clnoise} we address a pair of classical vdP oscillators under noise, in order to quantify 
how much the classical noise prevents the oscillation death as compared with the quantum noise. 
Section~\ref{collective} discusses collective oscillation collapse within the mean-field theory for 
a large number of oscillators as well as the exact solution of the master equation for small number of 
oscillators. Section~\ref{cd} concludes our results, and we remark on the possible implementation of vdP oscillators.

\section{A pair of classical vdP oscillators}\label{N2cl}

We introduce a dissipative coupling of the strength $V$ between a pair of classical vdP oscillators, 
\begin{eqnarray}
\label{eq:amp2}
\dot{\alpha}_{j}=-i\omega_{j}\alpha_{j}+\frac{G}{2}\alpha_{j}-\kappa |\alpha_{j}|^2 \alpha_{j} + \frac{V}{2} (\alpha_{j'}-\alpha_{j}),
\end{eqnarray}
where $\alpha_{j}$, $\omega_{j}$ $(j=1,2)$ are the amplitude and intrinsic frequency of the $j$th oscillator, respectively,
and $j' \ne j$. 
We have assumed that the radius of each limit cycle is equal for two oscillators in the absence of the coupling.

The coupling leads to intriguing amplitude and phase dynamics. 
When the coupling $V$ and frequency mismatch $\Delta = \omega_1-\omega_2$ are increased, 
the self-oscillation is frozen and 
the rest state $|\alpha_{1}|=|\alpha_{2}|=0$ substitutes a stable fixed point~\cite{KBE, Aronson:1990uq}. 
The linear stability analysis predicts that the rest state is stable in the regime $G<V<(\Delta^2+G^2)/(2G)$~\cite{Aronson:1990uq}. 
These boundaries are drawn in Fig.~\ref{fig:OQ-Quantum} with the solid line and curve. 
We emphasize that the rest state is always unstable without coupling, and 
this oscillator death is thus genuinely the consequence of interaction.

We also remark on the classical phase dynamics derived from Eq.~(\ref{eq:amp2}). 
The relative phase between two oscillators is locked in the regime $V > |\Delta|$, the so-called ``Arnold tongue"~\cite{Pikovsky}, where a pair of vdP oscillators synchronize. 
The boundary $V = |\Delta|$ (dashed line in Fig.~\ref{fig:OQ-Quantum}) is, however, derived by neglecting the amplitude variation associated with the coupling. 
The numerical solution of Eq.~(\ref{eq:amp2}) reveals that the synchronized motion in the steady state is seen only in the regime 
$V > (\Delta^2 +G^2)/(2G)$.

\section{A pair of quantum vdP oscillators}\label{N2qm}

The quantum model corresponding to Eq.~(\ref{eq:amp2}) is described by the master equation~\cite{Lee:2014dq, Walter:2014cs}, 
\begin{eqnarray}
\label{eq:two-vdp-me}
\begin{split}
\dot{\rho}=\sum_{j=1}^{2} ( -i[H_{j},\rho] +G\mathcal{D}[a_{j}^{\dagger}] \rho+\kappa \mathcal{D} [a_{j}^{2}] \rho ) \\
 +V\mathcal{D}[a_{1}-a_{2}] \rho,
\end{split}
\end{eqnarray}
where $\rho$ is the density matrix of two oscillators, $a_{j}$ the annihilation operator of a phonon of the $j$th oscillator, and $H_j = \omega_j a_j^{\dagger}a_j$. The dissipator is defined as $\mathcal{D}[\mathcal{O}]\rho=\mathcal{O} \rho \mathcal{O}^{\dagger}-(\mathcal{O}^{\dagger}\mathcal{O} \rho + \rho \mathcal{O}^{\dagger}\mathcal{O})/2$ with ${\cal O}$ being an arbitrary operator. 
Here and henceforth we set $\hbar=1$. 
The term including $V$ denotes the dissipative coupling between two oscillators. 
\begin{figure}[t]
\includegraphics[scale=0.55]{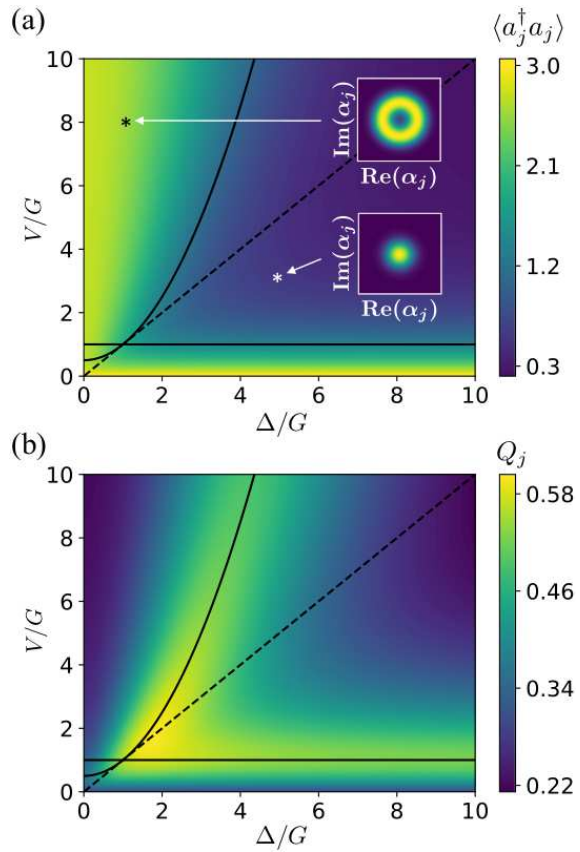}
\caption{(a)~Mean phonon number $\langle a_1^{\dagger}a_1 \rangle = \langle a_2^{\dagger}a_2 \rangle$ and  
(b)~Mandel $Q$ parameter $Q_1=Q_2$, 
in the steady state of the master equation~(\ref{eq:two-vdp-me}) for $\kappa/G = 0.2$. 
Solid line and curve denote the classical boundaries between the periodic synchronized motion and the oscillator death. 
Dashed line is the classical boundary of the phase synchronization derived by the neglect of the amplitude dynamics. 
Insets of (a) show the Wigner functions $W(\alpha_1)=W(\alpha_2)$ for $V/G=8,\ \Delta/G=1$ (upper panel) and for $V/G=3,\ \Delta/G=5$ (lower panel). Both insets show the same range $-5 \leqslant {\rm Re}(\alpha_j),\ {\rm Im}(\alpha_j) \leqslant 5$.}
\label{fig:OQ-Quantum}
\end{figure}

Recent studies on the quantum synchronization~\cite{Lee:2014dq, Walter:2014cs} pointed out that 
the phase synchronization occurs even in the deep quantum regime $\kappa/G \gg 1$ although the quantum fluctuations become significant when $\kappa/G \gtrsim 1$ as the radius of the limit cycle in the phase space decreases. 
As for the amplitude dynamics, on the other hand, the quantum ground state would not be anticipated because of the intrinsic quantum noise. The natural question here is, whether the oscillation collapse survives, or whether it is completely hindered by the quantum noise. If it survives, how small a mean phonon number is achievable, and what kind of measure characterizes this phenomenon? 
These are the central themes of this paper.

We solved the master equation (\ref{eq:two-vdp-me}) numerically~\cite{Johansson:2013gb} for a fixed value of $\kappa/G$, and show 
the mean phonon number of each oscillator $\langle a_1^{\dagger}a_1\rangle = \langle a_2^{\dagger}a_2\rangle $
in Fig.~\ref{fig:OQ-Quantum}(a) with respect to the coupling strength and the frequency mismatch, where $\langle {\cal O} \rangle = {\rm Tr}[{\cal O}\rho]$. 
In the absence of the coupling, the mean phonon number of individual oscillator is slightly larger than the squared classical amplitude, $\langle a_j^{\dagger}a_j\rangle\gtrsim |\alpha_{\rm ss}|^2=G/(2\kappa)$ for any value of $\kappa/G$. 
When they are coupled, however, the mean phonon number notably decreases in the regime where the classical oscillator death occurs, although the absolute ``death" of the classical oscillators is replaced by a moderate ``collapse" of the quantum oscillators.

The Wigner function is also shown in the insets of Fig.~\ref{fig:OQ-Quantum}(a). 
While the ring-shaped limit cycle is found in the classically synchronized regime, 
the oscillators settle into a vicinity of the quantum ground state for $G < V < (\Delta^2+G^2)/(2G)$ despite 
that the non-interacting vdP oscillators are not in the quantum regime. 
The results also indicate that the amplitude dynamics must be taken into account when we study the phase dynamics near the boundaries $V \simeq (\Delta^2+G^2)/(2G)$.

In order to characterize the quantum-mechanical oscillation collapse in terms of phonon-number statistics, 
we introduce the Mandel $Q$ parameter  
$Q_j = \langle (n_j -\langle n_j\rangle)^2\rangle/\langle n_j\rangle -1$ for each oscillator in the steady state where $n_j = a_j^{\dagger}a_j$. 
As shown in Fig.~\ref{fig:OQ-Quantum}(b), it increases near the classical 
boundaries of oscillation death. 
This noticeable increase in the number fluctuations indicates the quantum vdP oscillators drastically 
change their motional states between the periodic motion and the oscillation collapse. 
We note that $Q_j$ is 
always positive, consistent with the fact that the Wigner function of the steady state is always positive.

\section{Noisy classical model}\label{N2clnoise}

Since any quantum system involves intrinsic noise, 
the quantum dynamics should be compared with the classical dynamics with noise of which 
strength is the same as the quantum noise. 
Here we study the effects of such noise on the classical oscillation death outside the quantum regime, $\kappa/G \ll 1$. 
We determine the noise strength from the phase-space representation of 
the master equation~\cite{Walls}. The master equation~(\ref{eq:two-vdp-me}) is exactly rewritten in differential equations for the Wigner function $W(\alpha_1, \alpha_1^*, \alpha_2, \alpha_2^*)$ as, 
\begin{widetext}
\begin{eqnarray}
\dot{W}=\sum_{j=1}^2 
\left[
-\left(\frac{\partial}{\partial \alpha_j} \mu_{\alpha_j} + c.c\right)
+ \frac{1}{2}\left(\frac{\partial^2}{\partial \alpha_j \partial \alpha_j^*} D_{\alpha_j \alpha_j^*}
+\frac{\partial^2}{\partial \alpha_j \partial \alpha_{j'}^*} D_{\alpha_j \alpha_{j'}^*}
\right)
+ \frac{\kappa}{4}\left(\frac{\partial^3}{\partial \alpha_j^*\partial \alpha_j^2}\alpha_j + c.c.\right)
\right] W, \label{Weq}
\end{eqnarray}
\end{widetext}
where $\mu$ and $D$ are the elements of the drift vector and the diffusion matrix, respectively. 
The third-derivative terms in the Eq.~(\ref{Weq}) can be ignored when $\kappa$ is 
the smallest quantity, which is the situation we now consider. 
Under this circumstance, Eq.~(\ref{Weq}) reduces to the Fokker-Planck equation, 
and the equivalent classical stochastic differential equations are obtained (see the Appendix). 
We thus compare the quantum dynamics obeying the master equation, 
and the classical dynamics obeying the stochastic equation with a noise strength equal to that of the quantum system.

\begin{figure}[h]
\includegraphics[scale=0.55]{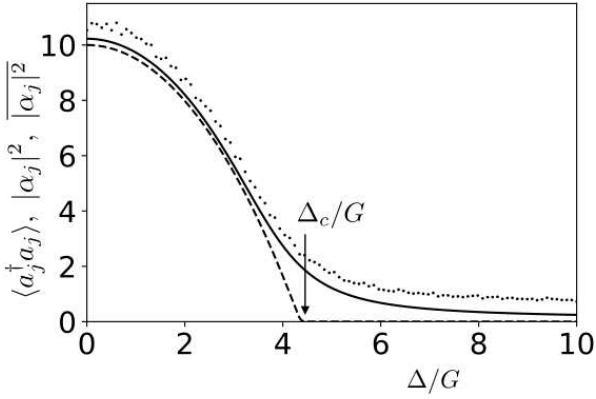}
\caption{
The mean phonon number of the individual quantum vdP oscillator (solid curve), 
the squared amplitude of the noiseless classical oscillator (dashed), and 
the average of the squared amplitude of the noisy classical oscillator (dots),
in the steady state for fixed $V/G=10$ and $\kappa/G=0.05$. 
}
\label{fig:qf-wn}
\end{figure}

Figure~\ref{fig:qf-wn} shows the ensemble average $\overline{|\alpha_j|^2}$ of squared steady-state amplitude of individual oscillators over 1000 runs of time evolution starting from independent initial states obtained from the classical stochastic equations (dots), as well as the squared amplitude $|\alpha_j|^2$ obtained from the classical amplitude equation without noise (dashed), and the steady-state phonon number $\braket{a_j^{\dagger}a_j}$ obtained from the master equation (solid curve) for a fixed value of the coupling strength. 
In the classical noiseless oscillators, the transition from the periodic motion to the oscillator death occurs at $\Delta_c = \sqrt{2VG-G^2}$, and the steady state is the rest state for $|\Delta| > |\Delta_c|$, as indicated by the dashed curve. 
In the quantum oscillators, the tendency of the decrease in the steady-state phonon number agrees with the squared amplitude of the noiseless classical oscillators. 
The deviation is the largest in the vicinity of the classical critical point $\Delta_c$, again signaling the abrupt change in the stability of the motional state. 
In the noisy classical oscillators, $\overline{|\alpha_j|^2}$ no longer converges to zero either, 
and its value is always larger than $\langle a_j^{\dagger}a_j\rangle$. 
Furthermore, the effects of the noise are more considerable in the regime $|\Delta| > |\Delta_c|$, namely, the oscillator death is more hindered. 
We thus conclude that both the quantum and classical noises hinder the abrupt oscillation death.
Nonetheless, the oscillation collapse indeed occurs in the quantum vdP oscillators in view of the increase in the number fluctuations near the classical critical point, 
and the mean phonon number is always smaller than in the noisy classical vdP oscillators.

\section{Collective collapse}\label{collective}

We finally study the collective amplitude dynamics when 
many quantum vdP oscillators are globally coupled. 
The master equation for coupled $N$ vdP oscillators is given by 
\begin{eqnarray}
\label{eq:Nmeq}
\dot{\rho} = \sum_{j=1}^N (-i[\omega_j a_j^{\dagger}a_j, \rho] + G {\cal D}[a_j^{\dagger}]\rho + \kappa {\cal D}[a_j^2]\rho) \nonumber\\
   + \frac{V}{N}\sum_{j=1}^N\sum_{j'=1}^N {\!} '\ {\cal D}[a_j-a_{j'}] \rho, 
\end{eqnarray}
where $\sum'_{j'}$ means that the case of $j'=j$ is removed from the summation. 
We again assumed that the gain and loss rates of each uncoupled oscillator are identical, and that 
the frequency distribution is uniform in the interval $[-\Delta/2, \Delta/2]$:
\begin{eqnarray}
\label{eq:uni-distr}
\begin{split}
g(\omega) = \left\{ 
\begin{array}{l}
1/\Delta \hspace{30pt} \omega \in [-\Delta/2,\Delta/2]\\
0 \hspace{50pt} {\rm others}
\end{array}
 \right.
\end{split}. 
\end{eqnarray}

When the number of oscillators $N$ is large enough, we may factorize the many-body density matrix as $\rho \simeq \bigotimes_{j=1}^N \rho_j$. 
This corresponds to the mean-field approximation as the factorization results in the neglect of correlations between oscillators. 
The ansatz yields a set of master equations for $j=1,2,\dots, N$, 
\begin{eqnarray}\label{eq:mean-vdp-me}
\dot{\rho}_{j}\!\!\!&=&\!\!\! -i[\omega_{j}a_{j}^{\dagger}a_{j},\rho_{j}] +G\mathcal{D}[a_{j}^{\dagger}] \rho_{j}
+\kappa \mathcal{D} [a_{j}^{2}] \rho_{j}\nonumber\\ 
\!\!\!&+&\!\!\!\frac{2V (N-1)}{N}\mathcal{D}[a_{j}] \rho_{j}
+ V (A[a_{j}^{\dagger},\rho_{j}] - A^*[a_{j},\rho_{j}]), 
\end{eqnarray}
and we have defined 
\begin{eqnarray}
A= \frac{1}{N}\sum_{j'=1}^N{\!}'\braket{a_{j'}}_j,\quad A^* = \frac{1}{N}\sum_{j'=1}^N{\!}'\braket{a^{\dagger}_{j'}}_j, 
\end{eqnarray} 
where $\langle \cdots \rangle_j$ denotes the average with respect to the one-body density matrix $\rho_j$. 
The quantity $A$ is identified as an order parameter of phase synchronization~\cite{Lee:2014dq}. 

\begin{figure}[t]
\includegraphics[scale=0.48]{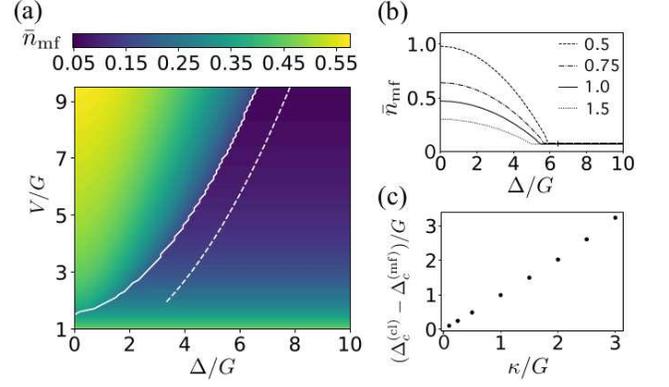}
\caption{(a) Mean phonon number per oscillator $\bar{n}_{\rm mf}$ in the steady state obtained from Eq.~(\ref{eq:mean-vdp-me}) for $N=100$ and $\kappa/G=1$. 
The value of $A$ is nonzero in the left region of the solid curve, and it is zero in the right region. 
Dashed curve corresponds to the classical boundary given by Eq.~(\ref{cpb}). 
(b) Mean phonon number $\bar{n}_{\rm mf}$ for a fixed $V/G=7$ for $\kappa/G=0.5,\ 0.75,\ 1.0,\ 1.5$. 
(c) The difference between the (noiseless) classical and quantum critical points $\Delta_c^{\rm (cl)}-\Delta_c^{\rm (mf)}$ 
versus $\kappa/G$ for $V/G=7$. 
}
\label{fig:ave-mean-p}
\end{figure}

We solve Eqs.~(\ref{eq:mean-vdp-me}) self-consistently, and 
show in Fig.~\ref{fig:ave-mean-p}~(a) the steady-state mean phonon number per oscillator 
$\bar{n}_{\rm mf} = \frac{1}{N}\sum_{j}\braket{a_{j}^{\dagger}a_{j}}\!_j$. 
When the frequency interval $\Delta$ increases with $V$ being fixed, 
the value of $A$ decreases and becomes zero at a certain value of $\Delta_c^{\rm (mf)}$, indicating the disappearance of the periodic synchronized motion. At the same time, $\bar{n}_{\rm mf}$ decreases and takes a constant value for $\Delta > \Delta_c^{\rm (mf)}$ as shown in Fig.~\ref{fig:ave-mean-p} (b). 
Outside the quantum regime $G \gtrsim \kappa$, we found that $\bar{n}_{\rm mf}$ behaves as $\bar{n}_{\rm mf} \simeq G/(2V-G)$.

The classical model corresponding to Eq.~(\ref{eq:mean-vdp-me}) is obtained by 
the replacement $\langle a_j \rangle_j$ with a $c$-number $\alpha_j$, 
as well as $\langle a_j^{\dagger}a_j^2\rangle_j \to |\alpha_j|^2\alpha_j$ in the equation of motion 
for $\langle a_j \rangle_j$ derived from Eq.~(\ref{eq:mean-vdp-me}). 
From the linear stability analysis of the ground state $|\alpha_j|=0$ for $j=1,2,\dots,N$, 
the phase transition from the periodic synchronized motion to the oscillator death is shown to occur~\cite{Ermentrout:1990} at
\begin{eqnarray}
\frac{\Delta_c^{\rm (cl)}}{2} {\rm{cot}} \left( \frac{\Delta_c^{\rm (cl)}}{2}/V_c^{\rm (cl)} \right)+ \frac{G}{2} - V_c^{\rm (cl)} = 0.\label{cpb}
\end{eqnarray}
This boundary is drawn as the dashed curve in Fig.~\ref{fig:ave-mean-p}~(a). 
Figure.~\ref{fig:ave-mean-p}~(c) shows the difference between the classical and quantum critical points $\Delta_c^{\rm (cl)}-\Delta_c^{\rm (mf)}$ as a function of $\kappa$ for a fixed value of $V$. 
As $\kappa/G$ decreases, the critical point $\Delta_c^{\rm (mf)}$ obtained from the mean-field theory approaches the classical value $\Delta_c^{\rm (cl)}$. On the other hand, when $\kappa/G$ increases and enters the quantum regime $\kappa/G \gtrsim 1$, 
the difference becomes larger and the oscillation collapse is seen even for smaller value of $\Delta/G$.

\subsection{Size scaling}

\begin{figure}[th]
\includegraphics[scale=0.55]{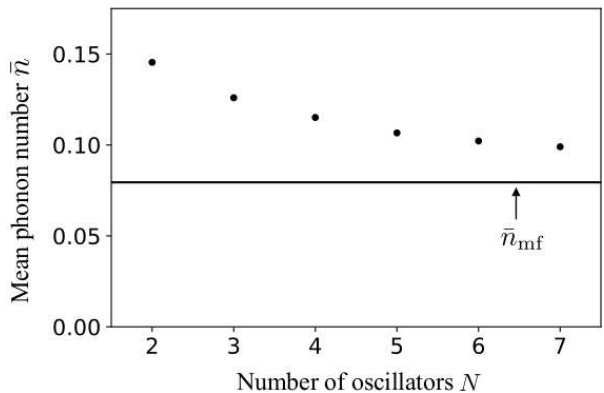}
\caption{Size effects on the oscillation collapse. 
Dots denote the mean phonon number per oscillator $\bar{n}$ obtained from Eq.~(\ref{eq:Nmeq}), and 
solid line denotes the mean-field result for $\kappa/G=100$. 
The coupling strength and the frequency mismatch are fixed as $V/G=5$ and $\Delta/G=10$ at which 
the mean-field theory predicts the oscillation collapse.
}
\label{fig:size-eff}
\end{figure}

The mean-field approximation is in general considered to be valid when the number of oscillators $N$ is large. 
We consider the opposite limit of a small number of oscillators in order to investigate size effects by exactly solving the many-body master equation~(\ref{eq:Nmeq}). The mean-phonon number per oscillator $\bar{n} = \frac{1}{N}\sum_{j}\braket{a_{j}^{\dagger}a_{j}}$, 
where the average is taken by the many-body density matrix $\rho$, is shown in Fig.~\ref{fig:size-eff} as a function of $N$. 
It takes the largest value for $N=2$ and decreases monotonically as the number of oscillators $N$ increases. It eventually approaches the result $\bar{n}_{\rm mf}$ of the mean-field theory as $(\bar{n}-\bar{n}_{\rm mf}) \sim N^{-1}$ within $2 \leqslant N \lesssim 7$ (the scaling is nontrivial for larger $N$). 
The oscillation collapse is thus collectively enhanced due to their global coupling: when the number of oscillators is increased, the mean phonon number per oscillator is lowered more. 

\section{Conclusions and discussion}\label{cd}

We have studied the oscillation collapse of coupled quantum vdP oscillators. 
The main results are summarized as follows: 
(i) large phonon-number fluctuations near the classical boundary between the periodic motion and oscillation death indicate that 
the periodic synchronized motion of a pair of quantum vdP also collapses into a vicinity of the quantum ground state when the frequency mismatch surpasses the critical value, 
(ii) a pair of quantum oscillators collapse closer into the ground state than the classical oscillators with the Gaussian white noise, and 
(iii)~we found collective enhancements of the quantum oscillation collapse in two ways: one is the widened oscillation-collapsed region as compared with the classical systems, and 
the other is the lowering of the mean phonon number for a larger number of oscillators. 

The vdP oscillators with global dissipative interaction may be realized, for instance, by placing dielectric membranes inside a Fabry-P\'erot cavity with a large out-coupling and by using multimodes of the cavity~\cite{Walter:2014cs}. The linear gain and the nonlinear loss terms in a single vdP oscillator are selectively controlled by placing the membrane in the vicinity of the node or antinode of the cavity mode. 
Driving the membrane with the blue- and red-detuned lasers, respectively, leads to the mechanical gain and loss, 
while the first blue-detuned laser is set on the single-phonon sideband and the second red-detuned laser is on the two-phonon sideband. The second laser induces the two-phonon loss. Furthermore, when the cavity out-coupling is large, the electromagnetic degree of freedom of the cavity is adiabatically eliminated, and the effective interactions of the form ${\cal D}[a_j-a_{j'}] \rho$ arise between membranes. 

The collective nature of the oscillation collapse may be utilized as an alternative mechanism for the cavity cooling 
of a number of mechanical oscillators of inhomogeneous frequencies. Furthermore, the substantial suppression of quantum fluctuations 
advocates the use of microscopic mechanical oscillators in noisy environments.

This work was supported by JSPS KAKENHI Grant No. JP16K05505.
 
\onecolumngrid
\section*{APPENDIX: DERIVATION OF CLASSICAL STOCHASTIC DIFFERENTIAL EQUATIONS}

We derive the classical stochastic differential  equations for a pair of vdP oscillators. 
The quantum master equation~(4) for $N$ vdP oscillators is rewritten as 
\begin{align}
\dot{\rho}
 =\sum_{j=1}^{N}  \left\{ -i[\omega_j a_j^{\dagger}a_j,\rho] +G\mathcal{D}[a_{j}^{\dagger}] \rho+\kappa \mathcal{D} [a_{j}^{2}] \rho +\frac{2V (N-1)}{N}\mathcal{D}[a_{j}] \rho\right\} 
-\frac{2V}{N}\sum_{j} \sum_{j'}{}^{\!'} \left(a_{j}\rho a_{j'}^{\dagger}-\frac{1}{2}a_{j'}^{\dagger}a_{j}\rho -\frac{1}{2}\rho a_{j'}^{\dagger}a_{j}\right). 
\end{align}
For $N=2$ we introduce a characteristic function~\cite{Walls}
\begin{align}
\chi(\bm{\beta})= {\rm{Tr}}[\hat{\mathscr{D}}(\beta_{1})\hat{\mathscr{D}}(\beta_{2})\rho],
\end{align}
where $\bm{\beta}= (\beta_1, \beta_1^*, \beta_2, \beta_2^*)$, and $\hat{\mathscr{D}}(\beta_{j})=e^{\beta_{j}\hat{a}_{j}^{\dagger}-\beta_{j}^*\hat{a}_{j}} \ (j=1,2)$\ is the displacement operator for the $j$th oscillator. 
According to the conversion rules from the products of the creation, annihilation, and density operators to the differential operators, 
the equation of motion for $\chi(\bm{\beta})$ is derived from the master equation as 
\begin{align}
\dot{\chi}(\bm{\beta})=&\sum_{j=1}^{2}\Bigg\{  \Big( i\omega_{j} \beta_{j}\frac{\partial}{\partial \beta_{j}} 
-i\omega_{j} \beta_{j}^* \frac{\partial}{\partial \beta_{j}^*}  \Big) 
+\frac{G}{2} \Big( -|\beta_{j}|^{2} + \beta_{j}\frac{\partial}{\partial \beta_{j}} +\beta_{j}^* \frac{\partial}{\partial \beta_{j}^*} \Big) \nonumber\\
& +\kappa\Bigg[ \beta_{j}\frac{\partial^3}{\partial \beta_{j}^* \partial \beta_{j}^2} +\beta_{j}^* \frac{\partial^3}{\partial \beta_{j} \partial {\beta_{j}^*}^2} + 2|\beta_{j}|^2 \frac{\partial^2}{\partial\beta_{j}\partial \beta_{j}^*} 
+ \frac{|\beta_{j}|^2}{4} \Big( \beta_{j}\frac{\partial}{\partial \beta_{j}} +\beta_{j}^* \frac{\partial}{\partial \beta_{j}^*} \Big)
+ \beta_{j}\frac{\partial}{\partial \beta_{j}} +\beta_{j}^* \frac{\partial}{\partial \beta_{j}^*} +|\beta_{j}|^2\Bigg] \nonumber\\
&\left. -\frac{V}{2} \left( |\beta_{j}|^{2} + \beta_{j}\frac{\partial}{\partial \beta_{j}} +\beta_{j}^* \frac{\partial}{\partial \beta_{j}^*} \right) \right. 
\left. + \frac{V}{2} \left( \beta_{j}\beta_{j'}^* + \beta_{j}\frac{\partial}{\partial \beta_{j'}} +\beta_{j'}^* \frac{\partial}{\partial \beta_{j}^*} \right) \right\} \chi(\bm{\beta}). \label{chieq}
\end{align}
Here and henceforth $j'$ denotes $j'=2, 1$ for $j=1, 2$, respectively. 
The Wigner function $W(\bm{\alpha})$ as a function of $\bm{\alpha} = (\alpha_1, \alpha_1^*, \alpha_2, \alpha_2^*)$ is the four-dimensional Fourier transformation of $\chi(\bm{\beta})$, 
\begin{eqnarray}
W(\bm{\alpha})=\int e^{\alpha_{1}\beta_{1}^*-\alpha_{1}^*\beta_{1}+\alpha_{2}\beta_{2}^*-\alpha_{2}^*\beta_{2}}
\chi(\bm{\beta}) d^2 \beta_{1} d^2 \beta_{2},
\end{eqnarray}
and the equation of motion for $W(\bm{\alpha})$ is obtained from the Fourier transformation of Eq.~(\ref{chieq}) 
and given by Eq.~(\ref{Weq}), with the drift and diffusion terms being given by  
\begin{eqnarray}
\mu_{\alpha_j} = \left(-i\omega_{j}+\frac{G}{2}-\kappa (|\alpha_j|^2-1)-\frac{V}{2}\right)\alpha_j + \frac{V}{2}\alpha_{j'}, 
\quad  
D_{\alpha_j \alpha_j^*}= G+ 2\kappa (2|\alpha_j|^2-1) + V,
\quad 
D_{\alpha_j \alpha_{j'}^*}= -V.
\end{eqnarray}
Equation~(\ref{Weq}) is exactly equivalent to the quantum master equation.

When the nonlinear loss rate $\kappa$ is small, the third-derivative terms in Eq.~(\ref{Weq}) may be ignored. 
This classical approximation~\cite{Lee:2013go} yields the Fokker-Planck equation, 
\begin{align}
\dot{W}(\bm{\alpha})=\sum_{j=1}^2 
\left[
-\left(\frac{\partial}{\partial \alpha_{j}} \mu_{\alpha_j} + c.c\right)
+ \frac{1}{2}\left(\frac{\partial^2}{\partial \alpha_j \partial \alpha_j^*} D_{\alpha_j \alpha_j^*}
+\frac{\partial^2}{\partial \alpha_j \partial \alpha_{j'}^*} D_{\alpha_j \alpha_{j'}^*}
\right)
\right] W (\bm{\alpha}) \label{cWeq}. 
\end{align}
For simplicity we move into the cartesian coordinates $\bm{X}=(x_1,y_1,x_2,y_2)$ by $\alpha_{j}=x_{j}+i y_{j}$, 
\begin{eqnarray}
\dot{W}(\bm{X})&=&\sum_{j=1}^2 
\left[
-\left(\frac{\partial}{\partial x_j} \mu_{x_j} + \frac{\partial}{\partial y_j} \mu_{y_j} \right)\right.\nonumber\\
&&\left.\qquad\quad  +\frac{1}{2}\left(
  \frac{\partial^2}{\partial x_j \partial x_{j}} D_{x_j x_j}
+\frac{\partial^2}{\partial y_j \partial y_{j}} D_{y_j y_j}
+\frac{\partial^2}{\partial x_j \partial x_{j'}} D_{x_j x_{j'}}
+\frac{\partial^2}{\partial y_j \partial y_{j'}} D_{y_j y_{j'}}
\right)
\right] W (\bm{X}) \label{cWeq}, 
\end{eqnarray}
where the drift vector ${\boldsymbol \mu}=(\mu_{x_1} \ \mu_{y_1} \ \mu_{x_2} \ \mu_{y_2})^T$ and the diffusion matrix ${\bf D}$ are, respectively, given by 
\begin{eqnarray}
&&\mu_{x_j}=\omega_{j}y_{j}+\left[\frac{G}{2}-\kappa (x_{j}^2+y_{j}^2-1)-\frac{V}{2}\right]x_j + \frac{V}{2}x_{j'}, \\
&&\mu_{y_j}=-\omega_{j}x_{j}+\left[\frac{G}{2}-\kappa (x_{j}^2+y_{j}^2-1)-\frac{V}{2}\right]y_j + \frac{V}{2}y_{j'},\\
&&{\bf D}=
\begin{pmatrix}
D_{x_1 x_1}&0&D_{x_1 x_2}&0\\
0&D_{y_1 y_1}&0&D_{y_1 y_2}\\
D_{x_2 x_1}&0&D_{x_2 x_2}&0&\\
0&D_{y_2 y_1}&0&D_{y_2 y_2}
\end{pmatrix}
=\frac{1}{2}
\begin{pmatrix}
\nu_1 & 0 & -V/2 & 0\\
0 & \nu_1 & 0 & -V/2\\
-V/2 & 0 & \nu_2 & 0 &\\
0 & -V/2 & 0 & \nu_2
\end{pmatrix},
\end{eqnarray}
and $\nu_1, \nu_2$ are defined by 
$\nu_{j}=G/2+\kappa [2(x_{j}^2+y_{j}^2) -1] +V/2$. 

The stochastic differential equation equivalent to Eq.~(\ref{cWeq}) is 
\begin{eqnarray}
\label{sde}
d {\bf X}= {\boldsymbol \mu} \ dt+ {\boldsymbol \sigma}\ d\bm{W}_t, 
\end{eqnarray}
where the noise strength ${\boldsymbol \sigma}$ and the diffusion matrix ${\bf D}$ is related as 
$\sum_{k=1}^{M} \sigma_{jk} \sigma_{kj'} = D_{jj'} $, and $d\bm{W}_t$ is the Wiener increment. 
Since the diffusion matrix ${\bf D}$ is symmetric, the elements of ${\boldsymbol \sigma}$ are analytically derived~\cite{Franck, Risken} as follows. 
The matrix ${\bf D}$ is diagonalized with the use of 
\begin{eqnarray}
{\bf U}=
&
\begin{pmatrix}
0&u_-&0&u_+\\
u_-&0&u_+&0\\
0&1&0&1\\
1&0&1&0
\end{pmatrix},\quad 
\displaystyle u_{\pm}=-\frac{\nu_1-\nu_2\pm \sqrt{(\nu_1-\nu_2)^2+V^2}}{V},
\end{eqnarray}
as ${\bf D}' = {\bf U}^{-1} {\bf D} {\bf U} = {\rm diag}(\lambda_-\ \lambda_-\ \lambda_+\ \lambda_+)$ where $\lambda_{\pm}$ is defined as 
$\lambda_{\pm}=
\frac{1}{4}\left( \nu_1+\nu_2\pm\sqrt{(\nu_1-\nu_2)^2+V^2} \right)$. 
The matrix $\bm{\sigma}= {\bf U}\sqrt{{\bf D}'}\ {\bf U}^{-1}$ is thus given by 
\begin{align}
{\boldsymbol \sigma}=\frac{1}{u_+-u_-}
\begin{pmatrix}
u_+  \sqrt{\lambda_{+}}-u_{-} \sqrt{\lambda_{-}} & 0 & \sqrt{\lambda_{+}}-\sqrt{\lambda_{-}}&0\\
0 & u_+  \sqrt{\lambda_{+}}-u_{-} \sqrt{\lambda_{-}} & 0 & \sqrt{\lambda_{+}}-\sqrt{\lambda_{-}}\\
\sqrt{\lambda_{+}}-\sqrt{\lambda_{-}} & 0 & u_+  \sqrt{\lambda_{-}}-u_{-} \sqrt{\lambda_{+}} & 0\\
0 & \sqrt{\lambda_{+}}-\sqrt{\lambda_{-}} & 0 & u_+  \sqrt{\lambda_{-}}-u_{-} \sqrt{\lambda_{+}}\\
\end{pmatrix}. 
\end{align}

\twocolumngrid


\end{document}